# Multi-view Consistency in UML

Alexander Knapp[1] and Till Mossakowski[2]

[1] Universität Augsburg, Germany
[2] Otto-von-Guericke Universität Magdeburg, Germany

**Abstract.** We study the question of consistency of multi-view models in UML and OCL. We first critically survey the large amount of literature that already exists. We find that only limited subsets of the UML/OCL have been covered so far and that consistency checks mostly only cover structural aspects, whereas only few methods also address behaviour. We also give a classification of different techniques for multi-view UML/OCL consistency: consistency rules, the system model approach, dynamic meta-modelling, universal logic, and heterogeneous transformation. Finally, we elaborate cornerstones of a comprehensive distributed semantics approach to consistency using OMG's Distributed Ontology, Model and Specification Language (DOL).

## 1 Introduction

Hartmut Ehrig was a researcher whose broad scope of interests ranged from category and automata theory through algebraic specifications and graph grammars to models of concurrency, and in all these fields he achieved fundamental results and contributed far-reaching and novel ideas. It is sad that such a great researcher passed away far too early after his retirement.

One of the many themes of Hartmut Ehrig's research has been the multi-viewpoint integration in the specification of complex systems [12]. We here address this problem more specifically in the context of the "Unified Modeling Language" (UML [34]). UML is a complex visual language featuring 14 different diagram types which may be complemented by textual annotations in the "Object Constraint Language" (OCL [33]); both languages are standardised by the Object Management Group (OMG). Already for UML 1.1 van Emde Boas observed "that UML is not a single language but a hybrid of several languages" [11] and Cook et al. [7] coined the notion of UML as "a family of languages". The multitude of diagram types and sub-languages offered by the UML/OCL allows the modeller to reduce the complexity of a model by specifying a system from different viewpoints: data, behaviour, interaction, component architecture, etc. Multi-view modelling and the necessity to integrate views devised from different viewpoints has been intensively discussed in the literature in general by Hartmut Ehrig et al. [12] and others [4], in the software architecture community [8, 21, 22], in the UML community [5], in the SysML community [32, 39], and also in other communities [2, 17, 35].

A central question in multi-view modelling is whether such a family of UML/OCL diagrams and annotations is still consistent, i.e., conjointly instantiable such that all views from all viewpoints are satisfied w.r.t. their (well-defined) semantics [c46][3]. This consistency problem has already been stated in the early UML days [5, 13, 14], and has been addressed quite intensively in the literature. In particular, several categorisations for partitioning the consistency problem have been designed: Engels et al. [c26] suggest to distinguish between horizontal (or intra-model) and vertical (or inter-model) consistency, i.e., whether the views are on the same level of abstraction; as well as syntactic (structural well-formedness of the abstract syntax) and semantic consistency (compatibility of behaviour). Mens et al. [29] focus more on the intention of sub-languages and give a classification into structural vs. behavioral diagrams and their use on the specification vs. instance level. Allaki et al. [1] combine these schemes into a typological frame of mono- vs. multi-diagram, specification vs. instance, and syntactic vs. semantic, and furthermore add a taxonomy of consistency problems in a terminological dimension, mentioning incompleteness, ambiguity, contradiction, incompatibility, and anomaly. From a verification perspective, Hilken et al. [18] present a list of structural and behavioural verification tasks for UML models considering besides consistency the categories of consequence, independence, executability, and reachability. A structural verification task considers a single (integrated) system state only, whereas a behavioural task pertains to a sequence of states. In contrast to [c46], here "[c]onsistency problems are structural problems and do not involve behaviour" [18, p. 122].

---

[3] References prefixed with a 'c' refer to the multi-view UML/OCL consistency bibliography assembled in a separate list.

The large number of approaches to multi-view consistency in the literature has also been reviewed and summarised [3, 20, 28, 40, 41, 42]. In particular, Torre et al. [40, 41] systematically survey existing consistency rules. They find that most rules are syntactic (88.21% in [40] and 81.89% according to the more comprehensive [41]), and that most of the rules are related to class diagrams (71.58%), sequence diagrams (47.37%), and state machine diagrams (42.11%). Moreover, they deplore that "it appears that researchers tend to discuss very similar consistency rules, over and over again", and conclude that "much more work is needed to develop consistency rules for all 14 UML diagrams, in all dimensions of consistency (e.g., semantic and syntactic on the one hand, horizontal, vertical and evolution on the other hand)" [40].

In this paper, we want to take up the challenge posed by Torre et al.: What is needed to develop a notion of multi-view consistency for a multitude of the (and ideally all) 14 UML diagrams and the OCL? We first give a comprehensive overview over the existing approaches to multi-view UML/OCL consistency in the literature both for UML 1 and UML 2 starting from the surveys mentioned above. We list which diagram types and sub-languages of UML/OCL are covered by each approach, which consistency technique it applies, whether it tackles structural or behavioural consistency, and which formalism and tool it uses. Our main contribution here is to point out and survey the variety of techniques to a grip on consistency for a heterogeneous, multi-view language like UML/OCL, ranging from syntactic consistency rules over an overarching, semantic system model to heterogeneous transformations. We find that structural consistency is considered more often by far and that either structural, syntactic consistency rules or an encoding into a system model or some universal logic prevails.

Purely syntactic approaches, however, do not help in ensuring behavioural consistency, and having to encode all UML/OCL sub-languages into a single semantic model or formalism, somewhat neglects the inherent heterogeneity of this multi-view language and almost necessarily leads to a certain unnaturalness. More often than not, it would be preferrable to represent the structural and behavioural semantics of a sub-language in an appropriate semantic domain of its own and only to relate these sub-languages and their semantic domains by translations. We therefore outline, extending our previous work on a truly heterogeneous approach to UML/OCL semantics [c11, c42], a consistency approach based on distributed heterogeneity in the "Distributed Ontology, Model and Specification Language" (DOL), also standardised by the OMG.

## 2  Review of Approaches to Multi-view Consistency in UML

Tables 1 and 2 contain an overview of existing approaches to multi-view consistency both for UML/OCL 1 (up to 2004) and UML/OCL 2 (starting in 2005), roughly ordered by their date of appearance. While the literature on UML abounds, for our topic, our literature review is sound and also comprehensive. Our starting points were the surveys [3, 20, 28, 41, 42], the information of which we aligned, adapted, and extended by search queries on the Internet and personal experiences.

| Reference | CD | OD | SM | ID | AD | OCL | cons. | class. | Form./Tool |
|---|---|---|---|---|---|---|---|---|---|
| * Egyed [c18, c19] | ◑ | ◑ | ◑ | ◑ | | | T | s | VIEWINTEGRA |
| * Große-Rhode [c32, c33] | | ◑ | ◑ | ◑ | | | S | b | transf. syst. |
| Reggio et al. [c66] | ◑ | | ◑ | | | | U | (s/b) | CASL-LTL |
| McUmber, Cheng [c56] | ◑ | | ◑ | | | | U | b | SPIN |
| * krtUML [c14] | ◑ | ◑ | ◑ | | | | S | s/b | symb. trans. syst. |
| Bernardi et al. [c7] | | | ◑ | ◑ | | | U | (b) | Petri nets |
| * xUML [c57] | ● | | ● | ◑ | ◑ | ◑ | S | (s) | Exec. UML |
| * Küster et al. [c23, c26] | ◑ | ◑ | ◑ | ◑ | | | U | b | CSP/FDR |
| * Hausmann et al. [c25] | | | ◑ | ◑ | | | D | b | Graph transf. |
| Spanoudakis, Kim [c70] | ◑ | | | | | | U | s | Dempster-Shafer |
| Litvak et al. [c51] | | | ● | ◑ | | | U | b | BVUML |
| Rasch, Wehrheim [c64] | ◑ | | ◑ | | | | U | s/b | Z, CSP/FDR |
| * Wirsing, Knapp [c75] | ◑ | | ◑ | | | | T | s/b | univ. alg. |
| Kyas et al. [c47] | ◑ | | ◑ | | | ◑ | U | s/b | PVS |
| van der Straeten [c71, c72] | ◑ | | ◑ | ◑ | | | U | s | Desc. Logic |
| Amálio et al. [c2] | ◑ | ◑ | ◑ | | | | U | s | Z |
| Kim, Carrington [c41] | ◑ | | ◑ | | | | U | s | Object-Z |
| Diethers, Huhn [c16] | | | ◑ | ◑ | | | U | b | UPPAAL |
| Yang et al. [c76] | ◑ | | | ● | | | U | s | rCOS |
| Yeung [c78] | ◑ | | ◑ | | | | U | b | CSP, B |

**Table 1.** Overview of UML/OCL 1 consistency approaches. CD means class diagrams, OD object diagrams, SM state machines, ID interaction diagrams, AD activity diagrams (as a special case of state machines), and OCL the Object Constraint Language. A ● means support for at least a substantial subset of the diagram/sub-language type, a ◑ indicates that the diagram/sub-language is supported but only for a limited subset. The consistency technique of the approach is indicated by an S for "system model", D for "dynamic meta-modelling", U for "universal logic", or T for "heterogeneous transformation". An entry "s" in the class(ification) column means that structural, static consistency checks are supported, a "b" that behavioural, dynamic consistency is checked; if the indicator is set into parentheses, the consistency support is quite restricted. The last column shows the used formalisms and tools. An asterisk in front of the reference indicates that more information is given in Sect. 2.1.

| Reference | CD | OD | CMP | CSD | SM | ID | AD | OCL | cons. | class. | Form./Tool |
|---|---|---|---|---|---|---|---|---|---|---|---|
| Lam, Padget [c49] | | | | | ◑ | ◑ | | | U | b | π-calculus |
| Long et al. [c54] | ◑ | | | | ◑ | | | | U | s | rCOS |
| Lucas et al. [c55] | ◑ | | | | ◑ | | | | U | s | Maude |
| Okalas et al. [c61] | ◑ | | | | ◑ | | | | U | s/b | B |
| Rasch, Wehrheim [c65] | ◑ | | ◑ | | ◑ | ◑ | | | U | s/b | Z, CSP/FDR |
| Wang et al. [c74] | ◑ | | | | ◑ | ◑ | | | U | b | LTSA |
| Bellur, Vallieswaran [c6] | ◑ | | ◑ | | ◑ | ◑ | | | U | s | meta-model |
| Li et al. [c50, c53] | ◑ | | ● | | ◑ | ● | | | U | s/(b) | UTP |
| O'Keefe [c62] | ◑ | | | | ◑ | ◑ | | | U | b | Dynamic Logic |
| Shinkawa [c68] | ◑ | | | | ◑ | ◑ | ◑ | | U | b | CPN |
| Yao, Shatz [c77] | ◑ | | | | ◑ | ◑ | | | U | b | Petri nets |
| Zhao et al. [c79] | ◑ | | | | ◑ | ◑ | | | U | b | SPIN |
| Anastasakis et al. [c3] | ● | | | | | | | ◑ | U | s/(b) | Alloy |
| * Gogolla et al. [c29] | ● | | ● | | ◑ | ◑ | | ● | U | s/(b) | USE |
| Knapp, Wuttke [c43] | ◑ | | | | ● | ● | | | U | b | Hugo/RT |
| Sapna, Mohanty [c67] | ◑ | | | | ◑ | ◑ | ◑ | | U | s | SQL |
| Brændshøi [c8] | ◑ | | | | ◑ | ◑ | | | U | b | impl. |
| Banerjee et al. [c4, c5] | ◑ | | | | | ● | | | U | b | Rhapsody/LTL |
| * Cengarle et al. [c11] | ● | | | | | ◑ | | ◑ | T | s/b | institutions |
| * Alanazi [c1] | | | | | ◑ | ◑ | | | U | (b) | impl. |
| Hammal [c34] | ● | | | | ◑ | ◑ | | | U | (b) | Petri nets |
| Laleau, Polack [c48] | ● | ● | | | ◑ | ◑ | | | U | s | meta-model |
| * Broy et al. [c9, c10] | ● | ● | | | ● | ● | | | S | s/b | set theory |
| * Kuske et al. [c44] | ● | ● | | | ◑ | ◑ | | | U | (s/b) | graph. transf. |
| * Grönniger [c31] | ● | ● | | | ◑ | ◑ | ◑ | | S | s/b | Isabelle/HOL |
| Nimiya et al. [c58] | ◑ | | | | ◑ | ◑ | | | U | b | Alloy |
| Khai [c39] | ◑ | | | | ◑ | ◑ | | | U | s | Prolog |
| Ober, Dragomir [c59] | | | ● | ● | ● | | | | U | s/b | OMEGA2 |
| Puczynski [c63] | ◑ | | | | ◑ | ◑ | | | U | s/b | impl. |
| Gerlinger et al. [c28] | ● | | | ● | | | | ● | U | s | Common Logic |
| El Miloudi et al. [c21, c22] | ● | | | | ◑ | ◑ | | | U | s | Z |
| Khan, Porres [c40] | ● | ● | | | ◑ | | | ◑ | U | s | Desc. Logic |
| * fUML [c60] | | | | | | | ● | | S | s/b | Common Logic |

**Table 2.** Overview of UML/OCL 2 consistency approaches. The abbreviations are as in Tab. 1 extended by CMP for component diagrams and CSD for composite structure diagrams. Activity diagrams (AD) have an independent semantics in UML 2. Protocol state machines are not equipped with a diagram type of their own in UML 2; still, [c29] considers them independently.

From the 14 different UML diagram types (structural: profile, class, composite structure, component, deployment, object, package; behavioural: activity, sequence, communication, interaction overview, timing, use case, state machine [34, p. 681]), we combined, as usual, the sequence, communication, interaction overview and timing diagram into the single type of interaction diagram for conciseness; and we omitted the profile, deployment, package, and use case diagram. Package diagrams provide a means for namespace modularisation and the package structure may most of the time be resolved statically using fully qualified names; still, also the meaning of packages and their relationships has been discussed [9, 38]. Use case diagrams, though besides class diagrams the most used diagram type of the UML [10, 26], convey rather little semantics on their own [19, 23]. Deployment diagrams, assigning software artefacts to system elements, also show quite limited semantic content (but cf. [30]). Finally, profile diagrams are used to define a domain-specific UML extension, thus every instance would add a viewpoint of its own.

Our survey thus covers 11 sub-languages of the UML/OCL family, where the entry for interactions condenses the information on four diagram sub-types. Since we aim at multi-view consistency involving several viewpoints, we do not list approaches here that only consider a single diagram type. In particular, we leave out the consistency of class diagrams, possibly accompanied by object diagrams (see [6] for an overview) or state machines [27, 36, 37]. Class diagrams, however, have been, in fact, the first instance of UML consistency investigations [13, 14, 24].

We now first review the variety of techniques enabling consistency checking in the listed multi-view approaches, and then report on our general observations and findings.

## 2.1 Consistency Techniques

The most immediate and direct approach to consistency checking of UML diagrams and models uses consistency rules, mostly on the concrete or abstract syntax. These rules extend the well-formedness rules of the UML specification given in OCL [c13]. Another option for such rules is to use other kinds of logics, like description logics [c40, c71, c72]. Many modelling tools incorporate their own rule sets [c17, c20, c52, c73]. The survey by Torre et al. [41] lists 116 consistency rules studied in the literature, where 95 are syntactic (structural).

Syntactic checks are indispensable in any approach to consistency, but they do not suffice to uncover the more intricate behavioural consistency problems, e.g., whether a network of state machines admits a trace specified by an interaction. Advanced consistency approaches thus have to develop and rely on a behavioural semantics of the UML/OCL diagrams and sub-languages of discourse. The degree of integration of these semantics varies considerably with the proposed approaches in the literature. Though the borders can not be always drawn with full accuracy, we suggest a categorisation w.r.t. the emphasis which is given to the semantic heterogeneity of UML/OCL. In the "system model" approach a uniform realisation frame is built, into which all sub-language aspects are encoded. The "dynamic meta-modelling" approach dispenses with the encoding, but enriches the meta-model, i.e., the abstract syntax of the UML, by semantic information. The "universal logic" approaches still use an encoding, though now to a uniform formalism. Finally, "heterogeneous transformations" approaches aim at employing families of translations for relating sub-languages.

*System Model.* The "system model" approach, best exemplified by Broy, Groenniger et al. [c9, c10, c31], builds on a uniform semantic basis for covering all aspects of state and state change present in any UML sub-language to be considered. By representing every facet of a model, expressed in various diagrams, in one and the same instance of the system model, static as well as dynamic checks can be performed. The "Executable UML" (xUML [c57]) as well as the "Foundational Subset of the UML" (fUML [c60]) use such system models for comprehensive and integrated execution.

For states, the system model in [c9, c10], contains a data store built from classes, their attributes, and the inheritance relationship as well as the instances; a control store consisting of operations and stacked method calls; and an event store holding also messages. For state changes, it comprises control-flow and event-based state transitions systems enriched by time. This system model, though with some modifications, e.g., specialising the event store to a message store, has later on been encoded in Isabelle/HOL and parts of UML class and object diagrams, state machines, and sequence diagrams, as well as a subset of OCL have been represented in this system model [c31]. With the help of the Isabelle prover then both static consistency checks, like whether an inheritance relationship is acyclic, and dynamic consistency checks, like whether a sequence diagrams is realisable by a state machine, can be done.

The manual effort to write down these checks and perform them in an interactive theorem prover seems quite substantial, however. Not to the least part, this is owed to the necessary complexity of the system model. Automation of various consistency checks has not been the primary goal of the approach. Still, the approach supports a certain degree of variability by exchanging sub-theories of the encoded system model, and other languages, a programming language, for instance, can be integrated [c31] if they can also be represented adequately in the system model.

The "krtUML" approach by Damm et al. [c14] uses symbolic transition systems as its system model for a comprehensive semantics. Their choice of class diagrams and state machines targets real-time systems. Consistency checks are not the main goal but behavioural consistency may be added on the basis of this system model. They stress that "[b]ecause all diagrams are only views on one and the same model, the attempts to give semantics for separated UML diagrams fail in producing the right semantics for the entire UML" (p. 94), though this valid observation somewhat neglects the possiblity to integrate the relations between the UML diagrams and sub-languages as done, e.g., in the heterogeneous transformation approaches.

*Dynamic Meta-modelling.* Inspired by attribute grammars that extend the abstract syntax tree of a (textual) language by synthesised and inherited attributes for semantic and contextual analysis, the "dynamic meta-modelling" approach by Hausmann, Engels, et al. [c12, c24, c25, c27, c35, c69] extends the abstract syntax of the UML, i.e., its meta-model, by semantic concepts on this very meta level. In fact, the UML meta-model shows several concepts that serve as links between the various sub-languages, like Event originating from, e.g., operation calls, used in state machines and activities for triggering behavioural effects or Message used in interactions for referring to operations and signals. By adding extra semantic concepts, the linkage between the sub-languages can be enhanced and, in particular, lifted to the behavioural, dynamic level. For example, the UML meta-class StateMachine is extended by a new meta-class EventPool for holding instances of the already existing meta-class Event that the instance of StateMachine then can react to; or the meta-class ControlFlow of activities is extended by a new meta-class ControlToken representing the possiblity that a control flow activity edge may carry a control token. Using the extensions, an operational semantics based on the extended meta-model and thus covering several UML sub-languages in concert is defined using (typed) graph transformations with negative application conditions, most prominently in the GROOVE tool [c35, c69] applying state space exploration. The graph transformation rules are separated into local small-step rules and transactional big-step rules that call the local rules.

This intriguing idea, combining attribute grammars and structural operational semantics, has mainly been applied to activities [c35] and to a limited degree to state machines [c69] and OCL [c12]. For consistency checks proper, the dynamic relation between sequence diagrams and state machines has been considered as an example, though without tool support [c25]. The overall design of dynamic meta-modelling somewhat resembles the "system model" approach as it builds a single domain of interpretation. By contrast, however, dynamic meta-modelling does not rely on an external semantic domain, but reuses the existing UML concepts and adds those features directly to the meta-model that are missing for behavioural interpretation. The use of a reference model for relating views that are then embedded and integrated into system model has already been advocated by Ehrig et al. [12]; the use of graph transformations on the meta-model level has also been used by Kuske et al. [c44]. Still, the complexity of the UML meta-model itself, let alone the necessary extensions, and respecting all semantic interconnections in local and global graph transformation rules present a major obstacle for the use of dynamic meta-modelling in comprehensive multi-view consistency checking.

*Universal Logic.* The system model approach builds a uniform semantic domain offering the necessary mechanism to interpret the different UML sub-languages and diagrams. By contrast, a "universal logic" approach does not rely on a single domain of interpretation, but just uses a uniform logical technique, like transition systems, for expressing the semantics of all UML diagrams to be checked for consistency. However, as in the system model approach, having to use a single encoding technique may sometimes yield unnecessary and unnatural complexity.

Große-Rhode [c32, c33] uses "transformation systems", i.e., extended labelled transition systems, where both states and transitions are labelled. The (control) states offer observations and synchronisation points, the transitions model the atomic steps of an entity and may be executed synchronously with other system parts. The semantics of class diagrams, state machines as well as their composition, and sequence diagrams are represented as classes of such transformation systems. Consistency can then be expressed by checking that the intersection of model classes (modulo some projections for adapting labels) are not empty. In a similar vein, though not as elaborated, the "superstate analysis" of Alanazi [c1] relies on nets of transition systems which allows to check the consistency of state machines and interactions.

The "Consistency Workbench" by Küster, Engels, et al. [c23, c26, c45] is based on a "partial translation of models into a formal language (called semantic domain) that provides a language and tool support to formulate and verify consistency conditions" [c23, p. 158]. In principle, the employed semantic domain is not fixed for all installments of the general approach and may vary with the consistency checks and the information extracted from the models by partial translation. The Consistency Workbench itself relies exclusively on the algebraic process language CSP and failure-divergence refinement (FDR). Still, it does not aim to construct an overarching system model, but is param-eterised in the consistency problem type. In this sense, it is bordering at an approach using heterogeneous transformation.

In the "film-stripping" approach by Gogolla et al. [c30, c36] a uniform technique for representing behavioural system evolution is used: System behaviour is captured by sequences of snapshots of system states, i.e., object diagrams, linked together by change information in particular recording how the objects evolve. Consistency checks could then be performed, e.g., in the USE tool [c29]. Even for automated analysis, like model checking, however, the general scaling of the technique without appropriate compression or abstraction of the snapshots remains unclear. The approach is complemented by model transformations from full-fledged UML to a simpler "base model" [c37, c38]: Complex modelling constructs, like association classes, are replaced with simpler modelling expressions, though possibly at the expense of having to use OCL. This technique is mainly exemplified by transformations on class diagrams and OCL itself, though, ultimately "[a]ll diagrams conjoined are transformed and combined into a base model" [c38, p. 60]. Thus, there are quite some similiarities with the system model approach.

*Heterogeneous Transformation.* Approaches based on "heterogeneous transformation" (coined by [c19]) focus on the several sub-languages and diagrams of the UML used in different forms at different development stages, from different viewpoints by different stakeholders [21, 22]. Such an approach has in particular been advocated by Derrick et al. [c15] for UML from their experiences with "Open Distributed Processing" (ODP) using Z and LOTOS, though not elaborated in detail. In their terminology a set of viewpoint specifcations is consistent "if there exists a specification that is a refinement of each of the viewpoint specifications with respect to the identified refinement relations and the correspondences between viewpoints" (p. 35).

Egyed's VIEWINTEGRA [c18, c19] defines transformations between the different UML diagram types on the very diagram level. In principle, all eleven diagram types of UML 1 could be covered, but only class diagrams, object diagrams, state machines, and interaction diagrams, i.e., sequence and collaboration diagrams (which, as in UML 2, are mere visual variants), are discussed in [c19]. The transformations are categorised into generalisation, e.g., object to class diagram; structuralisation, e.g., state machine to class diagram; translation, e.g., sequence into collaboration diagram; and abstraction, e.g., class diagram to class diagram. The last class of transformations, abstraction, is employed to relate diagrams and different development and refinement stages. Since the transformations map diagrams to diagrams, the supported consistency checks are structural; neither a static nor a dynamic semantics are provided. The classification of transformations is also used to reduce the number of necessary comparison transformations, which for eleven diagram types would otherwise be 55. However, when employing this design, not all transformations are possible any more, and a common denominator sometimes is needed, e.g., for comparing an object diagram with a state machine both are, perhaps somewhat arbitrarily, transformed into a class diagram.

An effort to formalise the relation of views and viewpoints on a semantic level is provided in [c75], where viewpoints are captured by a formal language category equipped with a model functor to a semantic domain category and views are language expressions in a viewpoint. Consistency is expressed by translations on the syntactic and semantic level; for semantic consistency a "viewpoint of comparison" has to be given.

Consistency checks, though only pair-wise, are exemplified for class diagrams, state machines, and sequence diagrams. The scheme by Cengarle et al. [c11] is similar, but uses the established theory of institutions as its foundation.

## 2.2 Observations and Results

Not all 53 approaches listed in Tabs. 1 and 2 are of the same quality w.r.t. elaboration and comprehensiveness with a kind of feasability study [c34, c58] and detailed accounts involving comprehensive semantics, tools, and larger case studies [c29, c43, c47, c59] as the ends. Tool support is similarly diverse and ranges from prototypical proprietary implementation over model checking and model finders to interactive theorem proving. In line with the survey results by Torre et al. [40], we also find that the diagram types most often covered are the class diagram (40 out of 53), the state machine diagram (44), and the interaction diagram (in one of its forms, 38); that according to our survey state machines have been considered more frequently than class diagrams may be accounted by our judicious choice of omitting single-view consistency approaches. Activity diagrams have rarely been integrated (5), but this may change with the broader adoption of fUML [c60]. The combination of state machines and interactions is considered most often (32). They are mainly considered in the context of class diagrams, rarely in combination with component diagrams and composite structure diagrams (3).

Among the approaches listed covering many different diagram types, the top four are the following ones:

– xUML [c57] covers five diagram types. However, for three of them, only a limited subset is covered, and only synactic, structural checks are provided.
– Gogolla et al. [c29] cover six diagram types, three of them to a substantial portion, and at least partially semantic, behavioural checks are provided.
– Broy et al. [c9, c10] define a comprehensive, though complex system model capturing substantial subsets of four diagram types of UML; however, tool support is not provided.
– Grönniger [c31], based on [c9, c10], integrates a fragment of the UML (partially covering class diagrams, object diagrams, interactions, state machines, and OCL) semantically by an encoding to the interactive theorem prover Isabelle/HOL.

All these either follow the system model or the universal logic approach. In accordance with the survey by Lucas et al. [28] we also find, that these encoding techniques (system model: 6, universal logic: 43) are currently by far prevailing for consistency. One major drawback of the encoding approaches, which may have prevented the further integration of other UML/OCL diagram types and sub-languages into them, is their lack of extensibility: A new viewpoint really extending the realm of expressivity inevitably calls for a considerable amount of work in expanding and adapting the already established semantic or logical domain.

## 3 Distributed Semantics for Multi-view Consistency

The previous section has shown that we are still quite far away from a comprehensive approach to multi-view consistency for UML. One reason is the complexity inherent in the diversity of UML models and their interaction. In [31], we have distinguished three different approaches to handle semantic heterogeneity:

– encoding of heterogeneous languages into some "universal" language. This approach is applied by both the system model, the dynamic meta-modelling and the univeral logic approaches described in Sect. 2;
– focused heterogeneity, i.e. languages are not per se encoded into some "universal" language, and complex models may involve parts written in different languages. Still, via translations, the end result is formulated in one language. This roughly corresponds to the heterogeneous transformation approach described in Sect. 2;
– distributed heterogeneity, i.e. truely decentralised networks of models formulated in different languages. This is a *heterogeneous transformation approach* that we propose to use.

Note that the existing heterogeneous transformation approaches are *not* distributed: they fall short in leveraging the translations to build up a distributed system of viewpoints. Therefore, they fail to handle the complexity of multi-view models in a comprehensive way. Such a comprehensive *and* semantic treatment is only possible with a decentralised approach, namely the distributed heterogeneous transformation approach. This means the provision of formal models for the involved UML/OCL diagram types and sub-languages that directly follow the intended semantics as specified in the UML/OCL standards. Here, the usual semi-formal language of mathematics is used, instead of casting the formal model into some predefined formal language. For a semantics distributed heterogeneity, these different formal models have to be linked, of course.

### 3.1 An institutional approach to distributed heterogeneous transformation

A useful meta notion for carrying out this formalisation is the notion of institution [15], an abstract formalisation of the notion of logical system. See [c56] for an early mentioning of institutions in the context of UML. The integration of different UML/OCL sub-languages (class diagrams, OCL, interactions) formalised as institutions has been started by Cengarle et al. [c11]. We have sketched the extension to further UML diagram types and a general vision in [c42].

A central feature of institutions is the provision of a notion of *realisation* of a model. Note that in institution-theoretic terms, this is called a model (in the sense of model theory) of a logical theory. However, this terminology can lead to confusion in the domain of model-driven engineering (MDE), where MDE models play the roles of logical theories. That is why we prefer the term "realisation" (of an MDE model) here. [18] speak of instantiation; however, this is tailored towards class diagrams. Realisations of models also exist for state machines (these are certain transition systems), sequence diagrams (there, they are trace sets), etc. They can differ greatly from institution to institution. The central point is that the notion of realisation provides a notion of *consistency*: a (single-view) model is consistent iff it has at least one realisation. Thus, the different notions of consistency from [18] can be captured by using different institutions. Moreover, these notions of realisation and consistency can be extended from the horizontal (or intra-model) case (i.e., living within one sub-language resp. institution) to the vertical (or inter-model) case of multi-view models, using so-called *networks*, see below.

We here describe general methods how to address the problem of multi-view consistency in this setting. Central tool is the Distributed Ontology, Model and Specification Language (DOL)[4], which recently has been adopted as a standard by the Object Management Group (OMG). DOL is not yet another modeling language, but rather a meta language for the specification of relation between different existing models. That is, UML diagrams can be referenced in DOL as-is, without the need of an encoding into some other language. The only need is to specify a formal semantics for the involved UML diagram types in the form of an institution. For UML class diagrams, this has been done in an informative appendix of the DOL standard itself.[5] For other UML diagram types, this has been partially done, see [c42] for an overview.

Translations between institutions can be formalised as so-called institution morphisms and comorphisms [16]. An institution morphism roughly corresponds to a projection from a "richer" to a "poorer" logic, expressing that the "richer" logic has some more features, which are forgotten by the morphism. The main purpose of the institution morphisms is the ability to express, e.g., that an interaction diagram and a state machine are compatible because they are expressed over the same class diagram. By contrast, institution comorphisms are often more complex. Roughly, a comorphism corresponds to an encoding of one logic into another one.

We now illustrate this framework with a few DOL examples.

### 3.2 ATM Example

In order to illustrate our approach to a heterogeneous institutions-based UML semantics in general and the institutions for UML state machines in particular, we use as a small example the design of a traditional automatic teller machine (ATM) connected to a bank. For simplicity, we only describe the handling of entering a card and a PIN with the ATM. After entering the card, one has three trials for entering the correct PIN (which is checked by the bank). After three unsuccessful trials the card is kept.

Let us assume that an institution morphism `sd2cd` from UML sequence diagrams to UML class diagrams has been defined. It extracts the classes, attributes, operations, signals etc. used in the sequence diagram and forms them into a class diagram. Given a sequence diagram `ATM_Bank_Interaction`, the DOL declaration for extracting its underlying class diagram is then

```
model ATM_Bank_Interaction_cd =
  ATM_Bank_Interaction hide along sd2cd
end
```

We now can express that a class diagram `User_Interface` refines to the sequence diagram `ATM_Bank_Interaction` in DOL:

```
refinement r1 =
    { User_Interface reveal ATM_Bank_Interaction_cd }
        refined to ATM_Bank_Interaction_cd
```

---

[4] http://dol-omg.org

[5] This is the first semantics of UML class diagrams that has been reviewed by co-designers of UML.

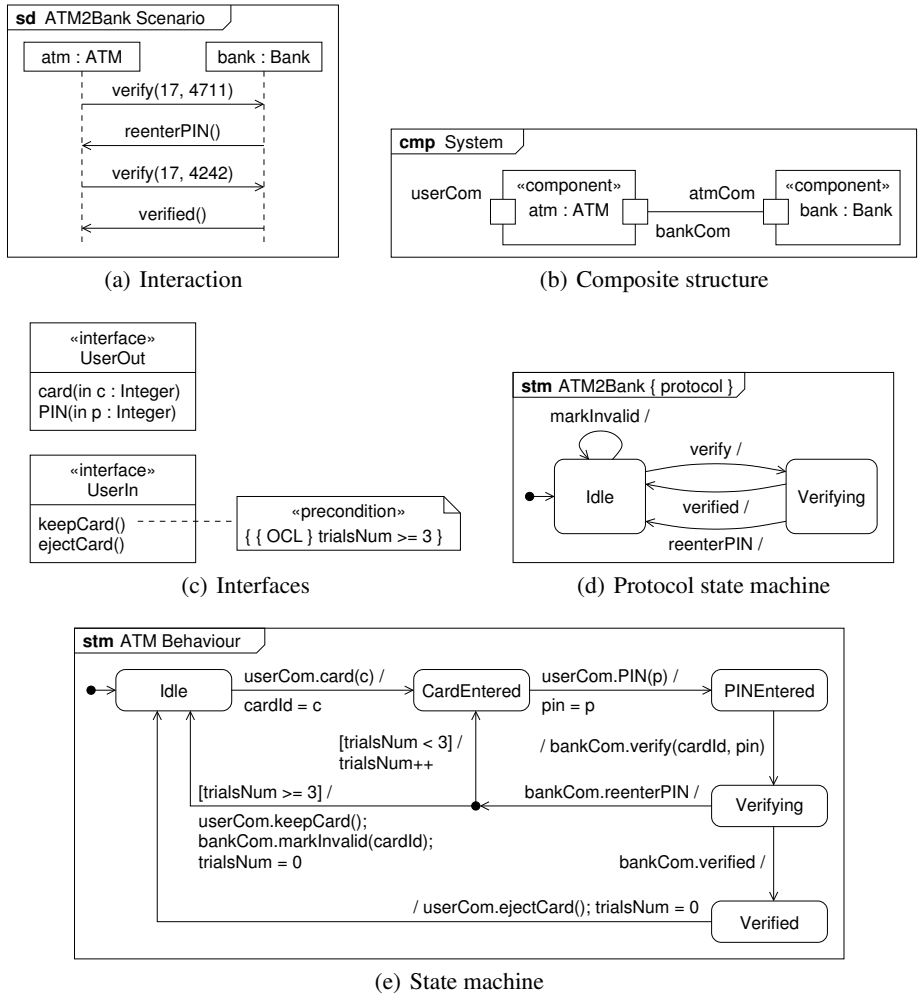

**Fig. 1.** ATM example

**end**

The semantics of such a refinement is inclusion of realisation. In this case, it means that each realisation of `ATM_Bank_Interaction_cd` is also a realisation of

```
{ User_Interface reveal ATM_Bank_Interaction_cd }
```

More specifically, this mean that each realisation of `ATM_Bank_Interaction` (which is a set of traces), when organised into a transition system of snapshots, is a transition system of the class diagram `User_Interface` (when the latter is restricted to those symbols that actually occur in `ATM_Bank_Interaction`).

Using a comorphism `cd2stm`, we can express that a state machine is built over some class diagram as follows:

```
model ATM_stm =
   User_Interface with translation cd2stm
then
   ATM_stm_definition
end
```

Now suppose that we have a second state machine (over the same class diagram, but typically not associated to the same class):

```
model Bank_stm =
   User_Interface with translation cd2stm
then
   Bank_stm_definition
end
```

and we have a system of these two state machines linked by a composite structure diagram within a component that we call `System`. This could be expressed as follows, using a comorphism `stm2cmp` linking state machines and composite structure diagrams:

```
model System =
   ATM_stm with translation stm2cmp with cid |-> atm
and
   Bank_stm with translation stm2cmp with cid |-> bank
then
   cmp
end
```

We now express that this system can realise the interactions expressed in sequence diagram `ATM_Bank_Interaction` as follows:

```
refinement r2 =
   ATM_Bank_Interaction refined to { System hide along cmp2sd }
end
```

Multi-view models can be expressed as so-called *networks* in DOL. A network consists of a number of component models (the views), plus some links (e.g. refinements) between these. So we obtain a network of the above UML diagrams and state its consistency as follows:

```
network N = %consistent
   User_Interface, ATM_stm, Bank_stm, System,
   ATM_Bank_Interaction, r1, r2
end
```

A realisation of the network consists of a family of realisations for each of the different components (views) of the network that is *compatible* along the links.[6] Consistency of the network means existence of at least one realisation. Note that consistency of a network is more than pairwise consistency of each pair of models involved.

---

[6] See [31] and the DOL standard at `http://dol-omg.org` for details.

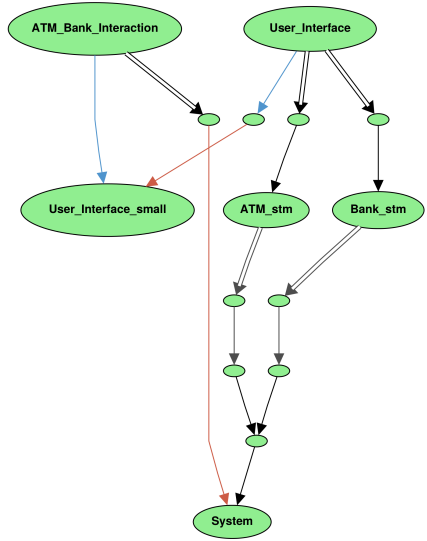

**Fig. 2.** Development graph for the ATM/Bank system

An important question is how to check consistency of such a heterogeneous multi-view network. One possibility follows the system model paradigm: using comorphisms, all involved models are translated into one formalism (e.g. first-order logic or Isabelle/HOL), and then consistency is checked there. This amounts to checking the consistency of a (homogeneous) logical theory, which can be done using standard realisation finders[7].

A second option is more decentralised and starts with separate realisations of all the involved models. Such realisations generally can be expressed in DOL: for example, for a class diagram, a realisation can be expressed as an object diagram. For a state machine, a so-called free DOL expression can be used to specify a particular DOL realisation. In order to check that such a family of realisations is a realisation of the network, pairwise compatibility of these realisations needs to be checked. Of course, this will generally fail if the invididual realisations have been randomly produced by a realisation finder. This means that the realisations need to be carefully designed in a manual way having in mind that they must compose into a realisation of the overall network.

This decentralised approach of finding a realisation is largely manual. More automation can be obtained using an incremental approach. Starting with class diagrams, OCL expressions and object diagrams, realisations for the static parts are obtained, e.g. using a realisation finder for first-order logic. In the next step composite structure diagrams and state machines are considered, and then sequence diagrams.

---

[7] Usually, these are called model finders, e.g. for first-order logic. However, in our terminology, a first-order model of a logical theory would be a realisation of (the translation) of an MDE model.

The use of realisation finders needed for consistency checking in all these approaches quickly comes to its limits when large models (or logical theories, after e.g. translation to first-order logic) are involved. In [25], we have developed an approach of decomposing large consistency problems into smaller ones, using so-called CASL architectural specifications.

## 4    Conclusion and Future work

UML/OCL is a language for multi-view and multi-viewpoint models, and the detection of view consistency at an early stage of the development is important for avoiding costly redesign. We have classified 53 existing approaches to UML/OCL multi-view consistency. Even the best approaches cover only five of the 14 UML diagram types, and most of these only partially. Moreover, a "universal logic" approach is predominant, where all UML/OCL diagram types and sub-languages are embedded into one system model or one logic. We have argued that this is not suitable for handling the involved complexity.

We propose a new approach to UML multi-view consistency, following a "heterogeneous transformation" paradigm. We use institutions for formalising the different UML diagram types and their semantics, and institution (co-)morphisms for formalising the transformations. Then UML multi-viewpoint models can be formalised as so-called networks in the OMG-standardised Distributed Ontology, Model, and Specification Language (DOL). This provides a framework where eventually all semantically relevant diagram types can be covered.

In order to use this framework for checking consistency of UML multi-viewpoint models, there is still a considerable way to go: while some UML diagram types have been formalised as institutions, this needs to be completed to a more comprehensive treatment of both diagram types and their features. Formalisation of transformations as institution (co-)morphisms has just started. In order to make this practically useful in connection with DOL, all these institutions and (co-)morphisms need to be integrated into the Heterogeneous Tool Set (Hets) and interfaced with suitable proof and model finding tools. Finally, suitable consistency strategies need to be developed and implemented.

Of course, writing down DOL expressions for large families of UML diagrams will be tedious. Hence, we aim at some graphical interface that can generate the needed DOL expressions automatically from a user's selection of those UML diagrams that should be interlinked to a network, plus a specification of the involved refinements. Such a specification of both networks and refinements adds the extra information to a given family of UML diagrams that is needed when checking multi-view consistency.

### Multi-view UML Consistency Approaches